# Use of 2G coated conductors for efficient shielding of DC magnetic fields.

J. F. Fagnard[1], M. Dirickx[1], G. A. Levin[2], P. N. Barnes[2], B. Vanderheyden[3], and P. Vanderbemden[3]

[1] SUPRATECS, CISS Department, Royal Military Academy, B-1000 Brussels, Belgium
[2] Propulsion Directorate, Air Force Research Laboratory, Wright-Patterson AFB, OH 45433 USA
[3] SUPRATECS, Department of Electrical Engineering and Computer Science, University of Liège, B-4000 Liège, Belgium

**Abstract**

This paper reports the results of an experimental investigation of the performance of two types of magnetic screens assembled from $YBa_2Cu_3O_{7-\delta}$ (YBCO) coated conductors. Since effective screening of the axial DC magnetic field requires the unimpeded flow of an azimuthal persistent current, we demonstrate a configuration of a screening shell made out of standard YBCO coated conductor capable to accomplish that. The screen allows the persistent current to flow in the predominantly azimuthal direction at a temperature of 77 K. The persistent screen, incorporating a single layer of superconducting film, can attenuate an external magnetic field of up to 5 mT by more than an order of magnitude. For comparison purposes, another type of screen which incorporates low critical temperature quasi-persistent joints was also built. The shielding technique we describe here appears to be especially promising for the realization of large scale high-$T_c$ superconducting screens.

## I. INTRODUCTION

Magnetic screening is of technological importance for a variety of applications requiring an ultra-low magnetic field environment, including biomedical [1], naval [2], or fundamental research instrumentation like SQUIDs [3] or cryogenic current comparators (CCC) [4] necessitating field attenuation larger than $10^6$ [5]. The traditional approach to shielding low-frequency magnetic fields is to use high permeability ferromagnetic materials such as permalloy or mu-metal [6]. At cryogenic temperatures (77 K and below), high temperature superconductors (HTS) often exhibit better low-frequency shielding performances than those of ferromagnets: e.g. magnetic field attenuations in excess of $10^5$ can be attained with Bi-2223 ceramics at 77 K [6]. With such an attenuation level, HTS are good candidates [5, 7] for replacing classical CCC based on low temperature superconductors [4].

Until recently, HTS shields have been made out of bulk materials [5, 8-10]. The significant progress in manufacturing the second generation (2G) coated conductors [11, 12] has opened a possibility that they can be used as a material for magnetic screens. A recent study [13] reported the use of YBCO coated conductors to attenuate AC magnetic fields at low frequencies (10 Hz – 100 Hz). A distinction should be made between "AC" and "DC" shielding. In the AC regime, the flux density $B(t) = B_0\sin(\omega t)$ at the sample surface induces alternating shielding currents that are limited by both the equivalent resistance $R$ and self-inductance $L$ of the current loop. The magnetic field can be attenuated if the resistance $R$ is *finite* provided the condition $R \ll \omega L$ is met. In the DC regime, however, an effective superconducting shield requires the existence of *persistent* current loops. In this paper, we show that loops made of 2G tapes can be candidates for replacing HTS bulks as efficient cylindrical DC magnetic shields.

## II. EXPERIMENT

The screening shell was assembled from eleven $12\times150$ mm$^2$ sections of coated conductor manufactured by SuperPower [14]. The average critical current ($I_c$) of these sections at 77 K is ~ 170 A. Similar to Ref. [15], along the center line of each 150 mm long section a 1 mm wide, 124 mm long slit was milled leaving a superconducting film in the form of a closed race track. Extended over a cylinder, such a section forms a superconducting closed loop. The magnetic shield, called Screen A, is assembled by stacking 11 sections, one on top of another, over a 60 mm diameter non-metallic holder to form a quasi-cylindrical shell. Screen A has an "eye-shaped" cross-section as shown in Fig. 1a. A preliminary characterization of the magnetic properties of each individual loop is carried out in order to place the best samples systematically in the central zone of the stack.

The properties of Screen A were compared with those of a more conventional solenoidal coil. Screen B is made out of a similar 2G YBCO 4 mm wide tape from American Superconductor ($I_c$ ~ 70 A at 77 K in self-field) [14]. The length of the solenoid is 80 mm (Fig. 1b). Its diameter (20 mm) was made intentionally smaller than that of Screen A. This allows us to insert Screen B into the sample chamber of a Physical Property Measurement System (Quantum Design) and to carry out measurements at temperatures down to 2 K. The ends of the solenoid can be either left open ("open" configuration) or connected by soldering an 80 mm long 2G tape ("closed" configuration). The superconducting critical temperature of indium used as a solder is 3.4 K.

The attenuation by a screen was determined as follows: the screen was zero-field cooled, then an axial magnetic field $B_{app}$ was applied and the magnetic flux density inside the shield, $B_{in}$, was measured using a high sensitivity Hall probe [9]. The shielding factor (*SF*) is defined as $SF = B_{app} / B_{in}$. The external field $B_{app}$ was ramped up from zero to the maximum value at a constant rate. The "DC" properties reported below correspond to a very low sweep rate (10 to 800 μT/s).

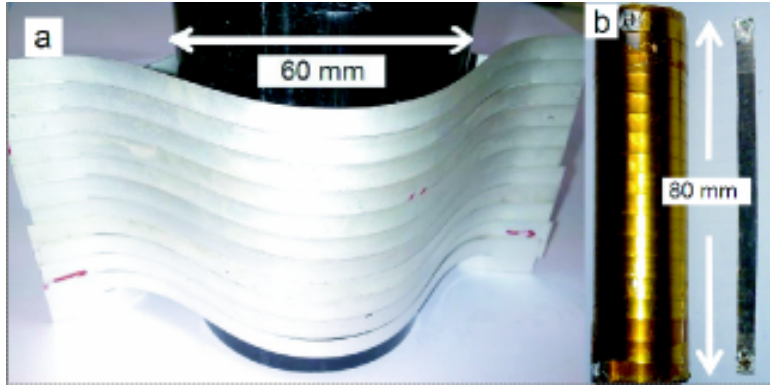

Figure 1. (a) Screen A – a screening shell assembled from 11 sections of 2G tape stacked together around a 60 mm diameter cylinder. (b) Screen B is an 18 turn solenoid which can be either open or short-circuited by a 80 mm long superconducting tape.

## III. RESULTS AND DISCUSSION

### III.1. Conventional solenoidal coil

First we examine the DC shielding factor of a conventional solenoidal coil (Screen B) under a 12 µT/s sweep rate. At $T = 77$ K (not shown), no shielding is observed in either configuration. Figure 2 shows $SF$ between 2.5 K and 4 K. The inset shows the corresponding $B_{in}$ vs. $B_{app}$ data. In the "open" configuration, $B_{in}$ equals the applied magnetic field at all temperatures. In the "closed" configuration, a magnetic shielding (i.e. $SF > 1$) is clearly observed at 2.5 K and 3 K. This is attributed to the superconducting transition of indium at 3.4 K. As a result, the contact resistance strongly decreases allowing much greater azimuthal screening current to flow through the solenoid. Note that the critical field of indium is 15 mT at 2.5 K and 7 mT at 3 K [16]. These values, indicated by arrows in Fig. 2, are consistent with the field amplitudes at which the shield effect disappears.

In coated conductors the interfacial resistance reported in the literature is of the order of 50 nΩ cm$^2$ [17-20]. In Screen B the contact area is 4×4 mm$^2$, so that a joint resistance ~100 - 300 nΩ seems to be a reasonable estimate. A solenoid like Screen B (self-inductance $L$ ~1 µH), "short-circuited" by a superconducting tape with two 100 nΩ joints, has a time constant $\tau$ ~ 5 s. For a given sweep rate d$B_{app}$/d$t$, the resistance of the joints will render magnetic shielding ineffective above a threshold induction $B_{lim}$ ~ $\tau$(d$B_{app}$/d$t$), i.e. ~ 0.06 mT for a sweep rate of 12 µT/s. Thus, the results of Fig. 2 provide direct experimental evidence that under small sweep rates ("DC" operation) an effective shield made of 2G superconductors requires a macroscopic persistent current.

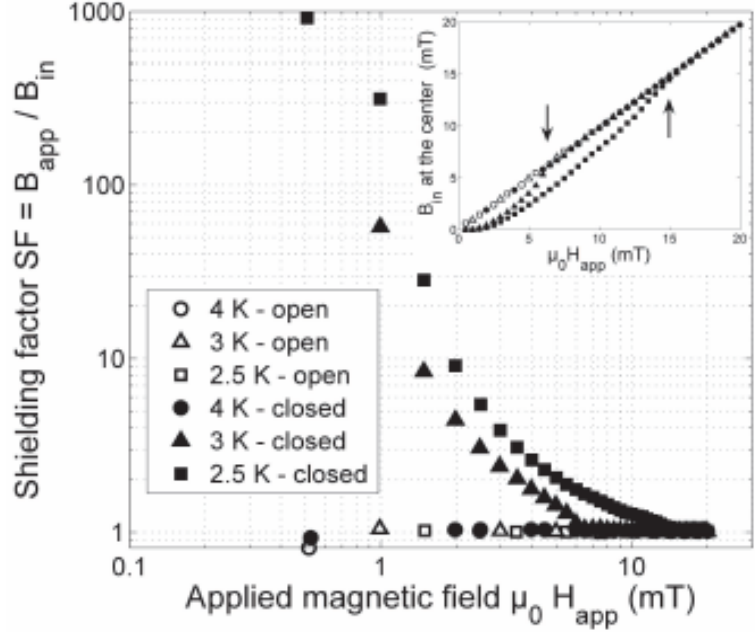

Figure 2. Shielding factor measured in the center of Screen B at 4 K, 3 K and 2.5 K. The open and solid symbols refer to the solenoid either in "open" or "closed" configuration, respectively. Inset: The dependence of $B_{in}$ vs applied field.

### III.2. Shield assembled with slit coated conductors

Let us now consider the effectiveness of the screening shell assembled from the slit conductors (Screen A, Fig. 1a). Figure 3 shows the shielding factor measured in the center of the coil (circles) at a 720 µT/s sweep rate; the inset shows a complete hysteresis curve with the external field $B_{app}$ ramped up to 20 mT and cycled subsequently between two symmetric values. As the applied magnetic field increases, $B_{in}$ is smaller than $B_{app}$ due to the shielding currents flowing along the tape loops. This leads to a shielding factor greater than 10 for applied fields up to 5 mT. The relatively low value of the shielding factor is mainly due to a small aspect ratio *l/D~1*, where *l is the length*, and *D the diameter* of the Screen A [6].

In order to estimate the influence of the finite size of the Screen A on its effectiveness, it is of interest to compare our measured shielding factor to the theoretical results expected for an ideal screen made of type-I superconductor of the same dimensions. Grohmann et al. [21] have shown that the internal magnetic field along the axis of a cylindrical semi-infinite type-I superconducting screen, close to the open ends, follows an exponential decaying function of the distance from the open end. Magnetic measurements and modelling have shown that this exponential law holds true for HTS tubes subjected to small magnetic fields [6]. A useful approximation of the shielding factor in the vicinity of the opening end is given by

$$SF \sim \exp\left(2C\frac{l/2-z}{D}\right), \qquad (1)$$

where $z$ is the elevation from the center of the cylinder and $C \approx 3.83$ is the first zero of the Bessel function of the first kind, $J_1(x)$. At the center of an ideal type-I superconducting tube, with roughly the same dimensions as those of Screen A ($l = 5.5$ cm and $D = 6$ cm), the shielding factor is $\sim 33.5$, a value which is close to the experimental values measured on Screen A at the lowest fields (circles in Fig. 3), e.g. $SF = 35$ at $B_{app} = 0.5$ mT. The agreement between our experimental results and the theoretical estimate for an ideal type-I screen thus confirms that the limitation of the shielding factor measured on Screen A is caused by the relatively small aspect ratio $l/D \sim 1$. For comparison, at the center of an ideal type-I superconducting tube, with roughly the same dimensions as those of Screen B ($l/D = 4$), the theoretical shielding factor would be $4.5 \; 10^6$, which is five orders of magnitude higher than for a shield whose aspect ratio equals 1.

The small value of the aspect ratio also has an impact on the threshold induction, $B_{lim}$, above which the shielding is ineffective. In the framework of the critical state model, the threshold induction $B_{lim}$ equals the full penetration field. Therefore the correcting factor, $CF$, of the threshold field for a type-II cylinder of finite length with respect to that of an infinite cylinder can be estimated from the correcting factor for the full penetration field; the analytical expression reads [22]:

$$CF = \frac{l}{2D}\ln\left[\frac{2D}{l}+\left(1+\left(\frac{2D}{l}\right)^2\right)^{1/2}\right]. \qquad (2)$$

If the "eye-shaped" cross-section is not taken into account, the correction for Screen A can be estimated to be $\sim 0.7$.

The inset of Fig. 3 shows the penetration of the magnetic field after the threshold value, $B_{lim}$. Above $B_{lim}$, the magnetic flux density inside the shield scales linearly with $B_{app}$. When the direction of $B_{app}$ is reversed, the internal magnetic flux density follows a hysteresis curve, yielding a remnant induction of 6.4 mT. The corresponding current circulating in each loop, by taking into account the cross-sectional shape and the finite length of the coil is estimated to be 42.5 A, which is smaller than the expected critical current of the slit tape. The reason is that the electromotive force ($S \times dB_{app}/dt$, where $S$ denotes the "eye" cross-section of a loop) gives rise to a smaller electric field $E$ than the usual threshold value $E_c = 1$ μV/cm used for the determination of $I_c$. A smaller current density is then expected because of the rather low $n$-values of $E = E_c (I/I_c)^n$ observed in coils made of 2G conductors [23].

Finally, $SF$ was also measured at several points along the $z$-axis. The results are qualitatively in agreement with Eq. (1). The measurements (Fig. 3) show that the shielding effect even manifests itself close to the top/botttom of the shell ($z = 3$ cm), where $SF$ values exceeding 3 are observed up to $B_{app} = 2$ mT.

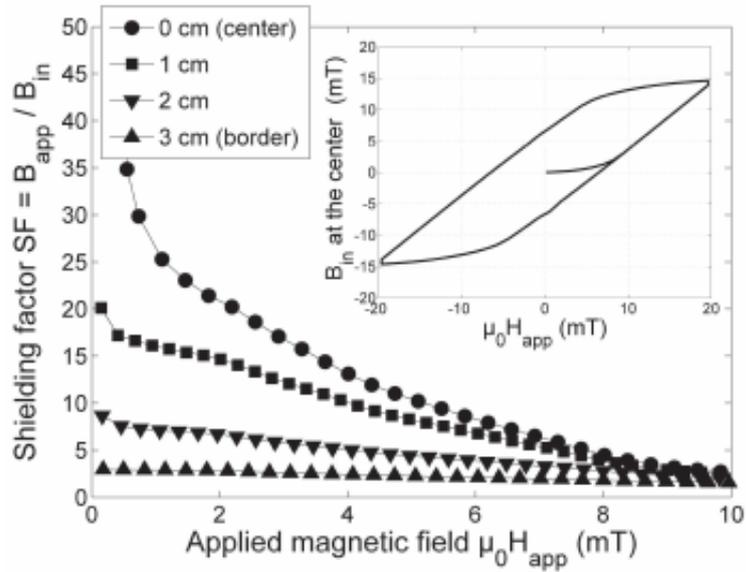

Figure 3. Shielding factor measured on the axis of Screen A at several elevations from the center. Inset: Hysteresis loop measured at the center of Screen A.

The measurements reported above demonstrate the potential of the screening shell assembled from the slit conductors for screening axial magnetic fields. It is of interest to comment on the ability of such shields for screening magnetic fields of any direction, and in particular transverse magnetic fields. The superconducting shielding of a transverse magnetic field is, in general, less efficient than the axial shielding because of demagnetizing effects. In the case of type-I superconductors subjected to a transverse magnetic field, it has been shown theoretically and experimentally that an exponential decay of the magnetic field occurs from the extremity of the cylinder similarly as in the axial configuration [21]. This decay is described by the same equation in both configurations (Eq. 1), where now $C \approx 1.84$ is lower than in the axial configuration. Similarly, our previous work on bulk HTS cylinders [9] has shown that both demagnetizing effects and a small finite aspect ratio lead to a $B_{lim}$ that is smaller in the transverse configuration than in the axial configuration. Another limitation arises from the fact that shielding currents can only flow in discrete areas limited by the width of the different tapes making the structure of the shield and not on the whole lateral surface of the shield. For these reasons, the stacked structure of Screen A is expected to provide shielding efficiency against a transverse magnetic field that is smaller than in the axial configuration. In order to increase the shielding efficiency in transverse configuration, the width of the tapes and the height of the screen should be as large as possible and a multilayer structure with overlaps should be preferred in order to reduce flux leakage between adjacent tapes.

## IV. CONCLUSION

In summary, the measurements reported above demonstrate a possibility to assemble a magnetic screen using state-of-the-art YBCO coated conductors. An effective screening of a low sweep rate ("DC") magnetic field is possible at temperatures as high as 77 K. The described method allows assembling a superconducting screen of practically unlimited volume with consistent properties using coated conductors instead of bulk superconductors. The circulating persistent current (and the screening factor) can be scaled up by stacking several slit coated conductors on top of each other before expanding such a loop over the coil former [15]. The neighboring sections of the screen can be made to overlap, preventing magnetic field leakage through the "magnetic cracks" between the adjacent tapes.

A potential large-scale application that can benefit from persistent magnetic shields made out of coated conductors is inductive fault current limiters (FCL) [24, 25]. These devices function similar to a transformer with superconducting short-circuited secondary winding. The superconducting winding (typically a cylinder) serves as a perfect magnetic shielding for the primary winding. This results in low residual impedance during normal operation. When the current in the primary winding exceeds a certain limit, the induced current in the superconducting shield also exceeds the critical current, leading to increased impedance.

For the inductive FCL, it seems likely that a large size superconducting shield assembled from coated conductors somewhat similar to the arrangement shown in Fig. 1 here and that in Ref. [15] will have advantage over a bulk YBCO in several respects – the consistency of the superconducting properties, potentially lower AC losses, as well as the lower cost and complexity of the manufacturing processes. More work on persistent magnetic shields made out of coated conductors is needed to clarify the scope of the problems to be solved and the scale of the potential advantages that may be associated with such shields.